\documentclass[preprint,showpacs,preprintnumbers,amsmath,amssymb]{revtex4}
\usepackage{graphicx}
\usepackage{psfig} 
\usepackage{changebar}
\usepackage[final,oztex]{epsfig}
\usepackage[applemac]{inputenc}
\psfigdriver{dvips}
\begin{document}
%%%%%%%%%%%%%%%%%%%%%%%%%%%%%%%%%%%%%%%%%%%%%%%%%%%%%%%%%%%%%%%%%%%%%%%%%
\renewcommand{\a}{\`{a}}
\renewcommand{\u}{\`{u}}
\newcommand{\e}{\`{e}}
\renewcommand{\o}{\`{o}}
\newcommand{\ii}{\`{\i}}
\def \nl{\par \noindent}
\newcommand{\E}{\`{E}}
\newcommand{\A}{\`{A}}
\renewcommand{\d}{{\rm d}}
\newcommand{\be}{\begin{equation}}
\newcommand{\ee}{\end{equation}}
\newcommand{\bea}{\begin{eqnarray}}
\newcommand{\eea}{\end{eqnarray}}
\renewcommand{\v}[1]{{\vec #1}} 
\newcommand{\dder}[1]{{{\rm d} \over {{\rm d} #1}}}
\newcommand{\pder}[1]{{\partial \over {\partial #1}}}
\newcommand{\sep}{\vspace {2truemm}}
\newcommand{\non}{\nonumber}
\pagenumbering{arabic}
%\renewcommand{\baselinestretch}{1.5}
%\psdraft 
%%%%%%%%%%%%%%%%%%%%%%%%%%%%%%%%%%%%%%%%%%%%%%%%%%%%%%%%%%%%%%%%%%%%%%%%%
%%%%%%%%%%%%%%%%%%%%%%%%%%%%%%%%%%%%%%%%%%%%%%%%%%%%%%%%%%%%%%%%%%%%%%%%%
%%%%%%%%%%%%%%%%%%%%%%%%%%%%%%%%%%%%%%%%%%%%%%%%%%%%%%%%%%%%%%%%%%%%%%%%%
%%%%%%%%%%%%%%%%%%%%%%%%%%%%%%%%%%%%%%%%%%%%%%%%%%%%%%%%%%%%%%%%%%%%%%%%%
%%%%%%%%%%%%%%%%%%%%%%%%%%%%%%%%%%%%%%%%%%%%%%%%%%%%%%%%%%%%%%%%%%%%%%%%%
\title{Dynamics of atoms in a time orbiting potential trap: consequences
of the classical description}
% \title{Atoms in a Magnetic Trap: some consequences of the Classical Model} 

\date{\today}

\author{Bruno Zambon}
\affiliation{INFM UdR di Pisa, Dipartimento di Fisica E. Fermi
Universit\`a di Pisa, Largo B. Pontecorvo 3, I-56127 Pisa, Italy.}
\author{Roberto Franzosi}
\email{franzosi@fi.infn.it}
\affiliation{C.N.I.S.M. UdR di Firenze and Dipartimento di Fisica,
Universit\`a degli Studi di Firenze, Via Sansone 1,
I-50019 Sesto Fiorentino, Italy.}

%%%%%%%%%%%%%%%%%%%%%%%%%%%%%%%%%%%%%%%%%%%%%%%%%%%%%%%%%
% ABSTRACT AND PACS
\begin{abstract}

The classical model that describes the motion of an atom in a
magnetic trap is solved in order to investigate the relationship between
the failure of the usual adiabatic approximation assumption and
the physical parameters of the trap. This allows to evaluate
the effect that reversing of the bias field rotation produces on the
vertical position of the atomic orbit, a displacement that is closely
related to the adiabatic character of the trap motion. The present
investigation has been motivated by a similar experimental test previously
carried out in the actual magnetic time orbiting potential trap.
We find that the non-adiabatic effects provided by the classical model
are extremely small. Thus, we conclude that the theoretical explanation
of the experimental measures, requires a quantum description of the
dynamics in magnetic traps.

\end{abstract}
%%%%%%%%%%%%%%%%%%%%%%%%%%%%%%%%%%%%%%%%%%%%%%%%%%%%%%%%%
%%%%%%%%%%%%%%%%%%%%%%%%%%%%%%%%%%%%%%%%%%%%%%%%%%%%%%%%%
\maketitle
%%%%%%%%%%%%%%%%%%%%%%%%%%%%%%%%%%%%%%%%%%%%%%%%%%%%%%%%%
%%%%%%%%%%%%%%%%%%%%%%%%%%%%%%%%%%%%%%%%%%%%%%%%%%%%%%%%%
% BODY OF PAPER
%%%%%%%%%%%%%%%%%%%%%%%%%%%%%%%%%%%%%%%%%%%%%%%%%%%%%%%%%
%%%%%%%%%%%%%%%%%%%%%%%%%%%%%%%%%%%%%%%%%%%%%%%%%%%%%%%%%

\section{Introduction}
In the realization of Bose-Einstein condensates (BECs) in
alkali atoms, the gas is confined and cooled in magnetic traps.
% Bose-Einstein condensation in alkali atoms, has been achieved
% by confining and cooling atoms in magnetic traps.
Among the different kinds of setups, in many experiments
have been used time-orbiting-potential (TOP) traps \cite{BECrew},
which are realized with two magnetic components: a static
inhomogeneous magnetic-field and an uniform rotating one.
The static inhomogeneous component originates a potential
well for the atoms, around a local minimum of the magnetic field.
The smaller this minimum is, the tighter the atomic confinement
is. Nevertheless, at zero magnetic field, atoms can leak from
the trapping potential as a consequence of Majorana transitions.
In order to avoid these atomic losses, in the TOP traps is added
the second magnetic component, that eliminates the zero of the
magnetic field at the trap center. The tight and stable confinement
of the atomic clouds, achieved in the TOP traps, makes these systems
well-suited for the Bose-Einstein condensation as well as
for quantum statistical/dynamical studies.
With respect to the theoretical investigation of the dynamics of atoms
in TOP traps, the {\em adiabatic approximation}, in which the atomic
magnetic moment is assumed to be constantly anti-aligned with respect
to the magnetic field, is usually assumed. A further possible mathematical
simplification, comes from the {\em time-average approximation}, in which
one assumes as potential for the atoms the time average of the true one.
Under these two approximations, the potential seen by an atom in a TOP
trap results harmonic and then the system Hamiltonian has trivial quantum
eigenstates.
However, such systems display a residual {\em micromotion} due to the fact
that the trapping potential is, indeed, time-dependent. This phenomenon has
been theoretically predicted in Ref. \cite{GS} on the base of a classical
model that prescinds from the adiabatic approximation, and it has been
experimentally observed in Refs. \cite{ARI,RIARI}. These latter papers,
have recently met the interest of several authors
\cite{pertot,bliokh,geursen,bouchoule,wheeler,ridinger}.
Other theoretical investigations on the quantum dynamics of atoms (condensates)
in TOP traps with the adiabatic approximation \cite{kuklov,minogin} 
(\cite{challis}), or without this approximation \cite{RF}, agree with the
experimental observations about this phenomenon.

A second surprising phenomenon has been observed in Ref. \cite{ARI}, in this
experiment, and for the first time, this team has been able to reveal an
{\em anomalous shift} in the vertical position of the atomic cloud, as the
rotating uniform magnetic field reverses its rotation. In the same paper it
has been hypothesized that the failure of the adiabatic approximation could be
at the origin of this phenomenon. To our knowledge, this anomalous shift
has never been explained from a theoretical point of view. Thus, the main
purpose of the present paper is to establish if, within a classical
description of the dynamics of an atom in a TOP trap, such phenomenon can
be explained. The {\em definite} outcome of the present paper is that the
non-adiabatic effects in the classical description of this system are extremely
small. Thus, such anomalous shift, cannot be explained within the classical scenario
commonly used to give a simple explanation of the mechanism behind cloud stability
in the magnetic traps. Consequently, we conclude that, the theoretical explanation
of this phenomenon, requires a quantum description of the dynamics of atoms in a
TOP trap.

This work is organized as follows. In section \ref{Model} and \ref{cyl}
we review some results and general ideas concerning the atomic motion
in a magnetic trap. In particular in section \ref{trapclassical} are introduced
the classical equations of motion describing the dynamics of an atom in a
magnetic trap, whereas section \ref{adiabaticapprox} is devoted to a general
discussion about the adiabatic approximation that simplifies drastically the
resolution of the equations of motion for this kind of systems.
In section \ref{cyl}, we derive several periodic orbits that take place in
a magnetic trap that has a cylindric symmetry around the vertical axis.
In section \ref{triaxial} we discuss the periodic motion in a
triaxial TOP trap similar to that used in the experimental investigation of
Ref. \cite{ARI} and we put a special emphasis on the reversing bias field rotation
test. Therefore, in Sec. \ref{zero} we derive the periodic orbit that takes place
in a triaxial trap for a very small asymmetry. In Sec. \ref{difference} we perform
a perturbative calculation that, by using the result of App. \ref{small},
allows us to evaluate the effect that the asymmetry produces on the dipole orientation
in space and the consequent reposition of the atomic trajectory along the vertical axis. Finally, in Sec \ref{difference}, we evaluate the consequent shift
in the vertical position of the atomic trajectory as the bias field reverses its
rotation.
In Section \ref{comments} we review some fundamental point concerning the
above experimental results, and the connected numerical simulation.
Thus we analyze the perspective emerging from the fact that the results provided
by the classical model do not agree with what is observed experimentally.

%%%%%%%%%%%%%%%%%%%%%%%%%%%%%%%%%%%%%%%%%%%%%%%%%%%%%%%%%
\section{Atomic motion in a magnetic trap}   \label{Model}

The confinement in space of a neutral particle carrying a magnetic moment,
can be obtained by means of the gradient magnetic-field forces experienced by
a magnetic dipole, in the presence of an inhomogeneous magnetic field.
This mechanism is at the basis of the atomic traps used by many experimental
groups working on the subject of Bose-Einstein condensation \cite{BECrew}.
In these traps, the atoms are first collected by means of laser forces in
a limited region of the space where such inhomogeneous magnetic field plays
its active role, and, afterwards, cooled down by an evaporative mechanism
so as to make the transition to a Bose-Einstein condensate possible.
So far the physics of the confinement by magnetic field has
been explained mainly by means of a set of equations in which the particle
is seen as a classical point-like magnetic dipole that obeys to the
classical equations of motion \cite{GS}.
Some quantum mechanical calculations exists for the
more simple configuration without the rotating bias field \cite{Bergeman}.
Nevertheless, this rotating field plays an important role in the TOP traps
and makes the physics of these systems more intriguing and complicated.
In fact, an important matter inherent this motion is the validity
of the adiabatic approximation. In the case where this approximation holds,
the magnetic dipole is constantly anti-aligned with respect to the magnetic
field and, thus, the magnetic energy of the atoms assumes the role of a space
confining potential. Existing quantum mechanical approaches
\cite{kuklov,minogin,challis} of atoms (or Bose-Einstein condensates) within a
magnetic trap, assumes from the start the fulfillment of the adiabatic
condition which is, actually, classical in nature. Indeed the
question concerning its range of validity has been tested experimentally in
\cite{ARI}. From these investigations it turns out that such an approximation
could become less stringent in the proximity of the marginal
stability configuration of a TOP  trap.  The critical test performed in
\cite{ARI} concerns the observation of an anomalous shift in the vertical
position of the atomic cloud as the bias radio-frequency field reverses
its rotation. The issue of adiabatic motion has also been theoretically
investigated for a symmetric trap, from which the triaxial trap used in
\cite{ARI} is obtained by setting the symmetry axis in the rotating bias
field horizontal plane. The dynamics in the cylindric trap has been studied
by means of a variational technique approach to the full quantum problem in
\cite{RF}. Whereas a semi-classical equations including quantum atomic
correlation has been used to investigate the more general problem
of the motion in a inhomogeneous magnetic field in \cite{yukalov}.

However, as said above, genuine quantum approaches to the study of the dynamics
of atoms in the TOP trap of Ref. \cite{ARI} does not exists.
These would have the merit of shedding light on the
condition of validity of the adiabatic approximation in the presence of the
spatial motion that a magnetic dipole confined in a magnetic trap undergoes.
These questions can also be related to the existence of a Berry's phases
phenomenology \cite{little}. While acknowledging
the need of a more detailed quantum investigation we will discuss here some
features of the magnetic confinement in a magnetic trap entirely from the classical 
point of view and in particular we will focus on the test of
the reversal of the bias field rotation. We believe that this has to be carried
out considering the importance and the intuitive support that this model provides in
explaining the trapping mechanism in the BEC experiments. Thus before of proceding with
a quantum computation it is necessary to assess clearly the scope of validity of the
already existing classical-model. This will be done here by checking the validity of
this model against the experimental results in \cite{ARI}. In particular when we will
refer generically to the experimental work or to the experimental measurements we will
mean experimental work and experimental measurements as done in Ref. \cite{ARI}.

\subsection{Trap classical equations} \label{trapclassical}
As a basic ingredient of a magnetic trap there is a magnetic field configuration
having a minimum of its intensity at a point in the space generally named trap
center. This is obtained with a static quadrupole magnetic field
whose components along the Cartesian axes are
\be 
b_x=b \, x \, , \quad 
b_y= b \, y \, , \quad 
b_z= -2\, b\, z  \, .
\label{inhB1}
\ee
Moreover, in order to prevent the spin flip transition that would occur when
the atom approaches the trap center, a rotating radio-frequency uniform
magnetic-field, usually known as bias field, is superimposed to the static
inhomogeneous one. In such a way the atom trajectory is kept quite far away
from this point. Let $B_b$ be the bias field intensity and $\omega$
its angular velocity in the $x-y$ plane. A different configuration for the
inhomogeneous field can be obtained by setting the symmetry axis to coincide
with the $x$ axis and by inverting the currents, this leads to
\be
b_x= 2\, b \, x \, , \quad 
b_y= - b \, y \, , \quad 
b_z= - b\, z \, , 
\label{inhB2}
\ee
which also provide a stable confinement.
We denote by  $\vec{B}$ the total magnetic field experienced by an atom which
is given by the sum of the inhomogeneous magnetic field and the bias rotating field.
The configuration where the inhomogeneous magnetic field is given by
Eqs. (\ref{inhB1}), is known as symmetric or cylindric TOP trap,
while the one corresponding to the configuration of Eqs. (\ref{inhB2}) is named
triaxial trap and is the one used in the experimental setup of Ref. \cite{ARI}.

Within the classical realm, the dynamics for an atom in a magnetic trap is given
by the equations of motion
\be
\begin{split}
m \ddot{\v{r}} &=  \v{F} = ( \v{\mu} \cdot \v{\nabla} ) \ \v{B}
+ m \, \v{g} \, , \\
\dot{\v{\mu}}& =  \gamma \, \v{\mu} \times \v{B} \, ,
\label{CEM}
\end{split}
\ee
where $\gamma$ connects the magnetic moment to the angular momentum of the
particle, i.e.  $\v{\mu}=g^\prime \frac{e}{2m_e}
\v{L}= \gamma \v{L} := \mu \v{n}$, here $m_e$ is the electron mass,
$g^\prime$ is the gyro-magnetic factor and $\v{n}$ is the unit vector
parallel to the magnetic moment.
In the discussion that follows we will use normalized units for the
quantities involved in the trap dynamics.
First of all we will assign a unit value to the period of the rotating
bias field and we adopt a length unit which coincide with the characteristic
length of the trap. Furthermore, in order to lift the constant in the
second of Eqs. (\ref{CEM}), we will include the $\gamma$ constant in
the magnetic field $\v{B}$, that will be measured in angular frequency units.
Consistently we will absorb the factor $1/\gamma$ in constant $\mu/m$.
For a typical TOP trap, we have a bias field frequency of the order of
$10$kHz and a characteristic length of the order of $1. \mu$m.
Thus the time unit will correspond to $T_0 = 10^{-4}$s, the length unit
to $L_0 = 10^{-6}$m, and the magnetic field intensity in the new unit,
will be obtained by multiplying the old value for the factor $\gamma T_0$.
E.g., a magnetic field of $10.$Gauss (order of magnitude
of the bias field used in \cite{ARI}) transforms to a field of about
$10^{-3} \frac{1.6 \times 10^{-19}}{9.1 \times 10^{-31}} 10^{-4} = 1.8
\times 10^4$, a field gradient $b$ of $100.$Gauss cm$^{-1}$ transform to
$\frac{10^{-2} \times 1.8 \times 10^{11} \times  10^{-4}}
{10^{-2} \times 10^{6}} = 18$. The parameter
$\mu/m$ of the first of Eqs. (\ref{SE}), in the new unit is obtained by
multiplying the original value for the factor $T_0/(L^2_0 \gamma)$.
Thus, by doing so for a trap, as the one of Ref. \cite{ARI},
operating with $^{87}$Rb atoms we have the following numerical values
$\frac{\mu}{m} = 0.036$, which is independent from the unit
chosen, $g=0.0981$, $\omega= 2 \pi$, $B_b \approx 10^4$ and a value of the
field gradient $b$ ranging from $0.1$ up to a value of $10.$
We will refer to these values in the following as the the standard trap parameters.
From Eqs. (\ref{CEM}) we finally obtain the normalized evolution equations
\be
\begin{split}
\ddot{\v{r}} &=  \frac{\mu}{m}( \v{n} \cdot \v{\nabla} ) \ \v{B}
+ \v{g} \, , \\
\dot{\v{n}}& =  \v{n} \times \v{B} \, ,
\label{SE}
\end{split}
\ee
where $\vec n$ is the unit vector corresponding to the atomic magnetic
dipole.

\subsection{Adiabatic approximation} \label{adiabaticapprox}
In the present section we summarize the conditions of validity of the
adiabatic approximation and its consequences about the dynamics of atoms in
magnetic traps.
The adiabatic approximation, that greatly simplifies the analysis of the atomic motion,
comes, pretty naturally, from an analysis of equations (\ref{SE}) and, in particular,
from the second one.
By indicating with $n_\|$ the projection of the vector $\vec n$ on the magnetic
field $\v{B}$, i.e. $n_\| = {\vec n} \cdot \frac{\vec B}{B}$, and with $\v{n}_\bot$ the
component of $\vec n$ perpendicular to this same field, that is $\v{n}_\bot =
\v{n} - (\v{n} \cdot \frac{\v{B}}{B}) \frac{\v{B}}{B}$, it is easy to show, by using
the second of Eqs. (\ref{SE}), that the following equality holds true
\be
{\dot n}_\parallel= {\vec n}_{\perp} \cdot  \frac{ \dot{\v{B}}}{B} \, .
\label{ADA}
\ee
From this equality it results clear that the validity of the adiabatic regime
is a consequence of the fast rotation of ${\vec  n}_\bot$ around $\v{B}$, that
takes place at an angular velocity of the order of $B$.
In fact, if $\frac{ \dot{\v{B}}} {B}$ has the same direction
of $\v{B}$, i.e. $\v{B}$ changes only in magnitude, the quantity 
$n_\parallel$ is exactly constant. If this is not the case, but
$\dot{\v{B}}$ is small, ${\dot n}_\parallel$ is also a small quantity with 
zero average because of the fast rotation of ${\vec  n}_{\perp}$ around $\v{B}$.
Thus, under the hypothesis of fast rotation of ${\vec  n}_{\perp}$ (that is $B >>1 $)
and small $\dot{\v{B}}$, we can conclude that ${ n}_\parallel$ is a quasi-constant of
motion and the magnetic energy of a particle in the trap results to be
\be
U=-\v{\mu} \cdot \v{B} = - n_\parallel \, \mu\, |\v{B}| \, .
\label{Umag}
\ee
Now the magnetic energy only depends on the projection of the magnetic moment
on the field $\v{B}$ and not from the orientation of the dipole in space, since
the other degrees of freedom have been frozen as a result of the adiabatic
approximation. Thus, a particle
anti-aligned with the field, or more in general a particle for which
$n_\|$ is negative, experiences a confining magnetic potential in the
proximity of a minimum of the magnetic field intensity.

In the case of the TOP traps, where the total magnetic field $\vec B$ is
time-dependent, it is more convenient to examine the torque equation (second
of (\ref{SE})) in a reference frame rotating around the $z$ axis together
the bias magnetic field. In this frame, the effective magnetic field experienced
by the particle is given by $\v{B}_{0}= \v{B} +\v{\omega}$, where
$\v{\omega}$ is directed along the $z$ axis. In the case of a
cylindric trap the latter will be a time-independent quantity, whereas for
the triaxial trap $\v{B}_{0}$ remains time-dependent.
Thus, in the case of the TOP traps, the adiabatic approximation has to be applied
in the reference frame rotating with the bias field and, therefore, it is the field
$\v{B}_{0}$ that plays the role of the field $\v{B}$ of Eqs. (\ref{ADA}) and
(\ref{Umag}).
Some questions related to the fact that a proper choice of the reference
frame really improves the analysis of a dynamical system when the adiabatic
approximation is to be used have been discussed in \cite{SCH}.
As a matter of fact it is out of doubt that in the TOP traps the frame
co-rotating with the bias field is the one that has such property.

In the following we will summarize briefly the peculiarities of the
motion in the symmetric and triaxial traps. In this last case, we will
calculate the shift of the vertical position of the atomic cloud, as
the bias field reverses its rotation. In the case of the symmetric trap
this shift is shown to be exactly $\Delta z= - \frac{\omega}{b}$. In fact,
a $-2 \omega$ change in the magnetic field $\v{B}_0$, is exactly
compensated by the $z$ component of the inhomogeneous field
in Eq.~(\ref{inhB1}), as the orbit
displaces itself of $\Delta z$  in the vertical position.
However, in the case of the triaxial trap, this argument does no more
hold true because its inhomogeneous field configuration does not
enjoy the rotational symmetry around the $z$ axis. This produces
a different dynamical evolution of the atomic magnetic moment, that
influences also the trap vertical position beyond the simple argument
given above.
The calculation of this contribution, will show how this shift is connected
to the failure of the adiabatic approximation, and to which extent this
mechanism can be understood within the limit of this classical model.

%%%%%%%%%%%%%%%%%%%%%%%%%%%%%%%%%%%%%%%%%%%%%%%%%%%%%%%%%
\section{Cylindric trap motion} 
\label{cyl}

\subsection{The periodic orbits} \label{periodicorbit}
\begin{figure}[ht]
\includegraphics[scale=0.4]{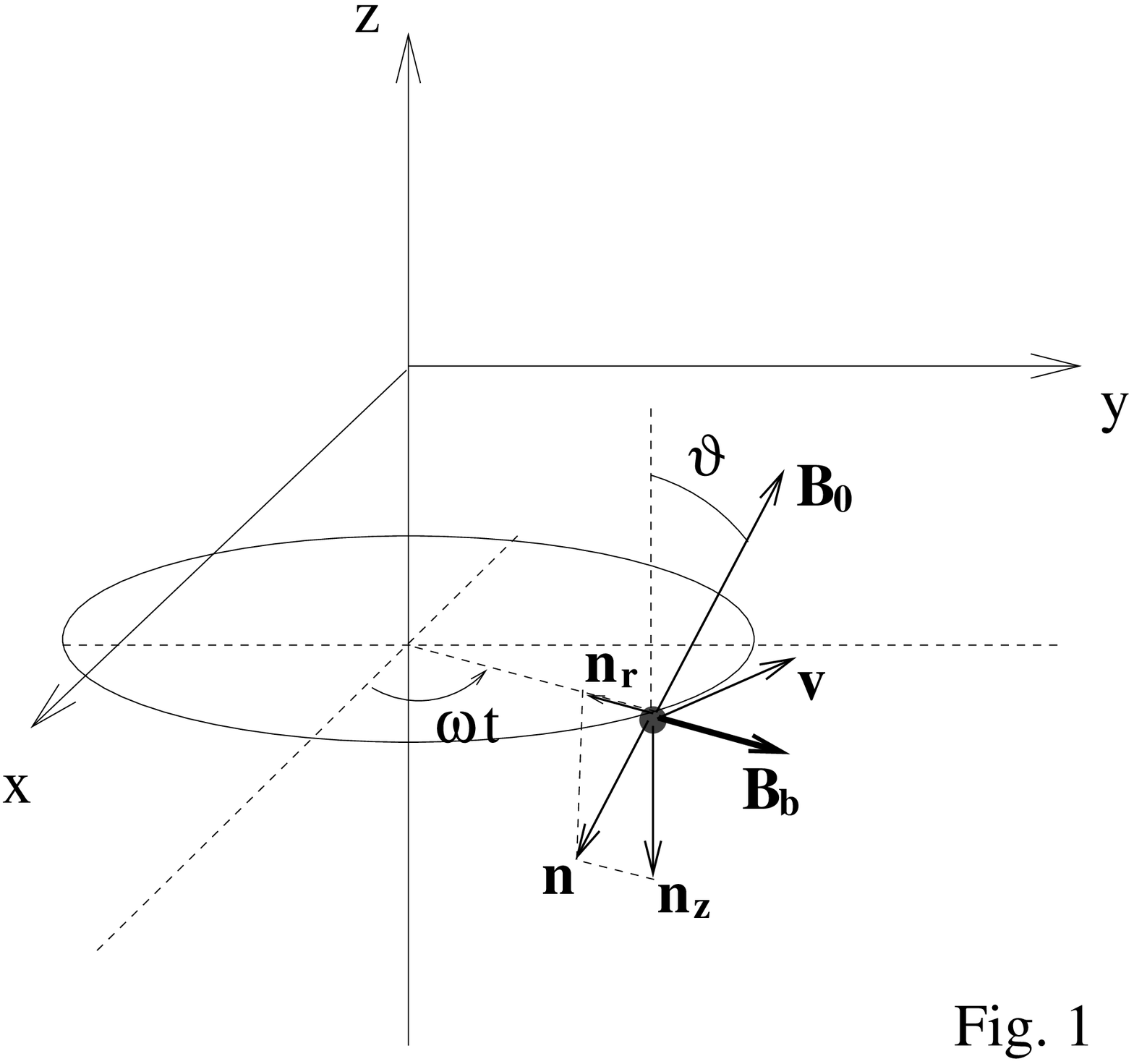}
\caption{The relative position of the magnetic moment, the bias field 
for a symmetrical trap configuration.}
\label{fig1}
\end{figure}

Stationary solutions of the classical equations (\ref{SE}) for the case of
a cylindrical trap, have been given in Ref. \cite{GS}. In order to derive
these solutions, and to find the stable orbits, Eqs. (\ref{SE}) are
transformed, and then solved in the reference frame rotating around the $z$
vertical axis which is opposite to the gravity
vector $\vec g$. This frame is the most convenient one because here the bias field
results to be time-independent as well as the fictitious
magnetic field $\v{B}_{0} = \v{B} + \v{\omega}$. In the latter expression
$\v{\omega}$ is directed along the $z$ axis, and $\v{B}$ includes both the bias
field and the inhomogeneous field. Remarkably, this latter has the cylindrical
symmetry around the $z$ axis. For the first of Eqs. (\ref{SE}), also the apparent
forces must be taken into account. There exists a stable solution in which the
magnetic dipole is anti-parallel to field $\v{B}_{0}$, this complying with the
torque equation, and the particle is at rest in this frame.
Thus the total force on the particle
must be zero. The projection of the equilibrium equation along the $z$ axis yield
$F_z=-  2\, \mu\, b \, \,n_z -m \,g=0$, from which we derive
\be
n_z =-\cos{\vartheta}= - \frac{ m \, g }{ 2\, b\,  \mu} \, .
\label{nz}
\ee
Let us consider now the equilibrium
in the $x-y$ plane. To fix the ideas we chose the rotating $x$ axis coincident
with the bias field (which has intensity $B_b$),
thus ${\vec B}_0 = (B_b + b x, b y, -2 bz + \omega)$ and
the gradient force has components in the $x-y$ plane given by $F_x = b \mu n_x$
and $F_y = b \mu n_y$. The balance with centrifugal force leads to
$n_x = - \frac{m \omega^2}{b \mu} x$ and $n_y = - \frac{m \omega^2}{b \mu}y$.
By combining this latter equation with the anti-alignment condition, that is
${\vec B}_0 \times {\vec n} = 0$, from $({\vec B}_0 \times {\vec n})_z = 0$
we obtain $y=0$. This means that the particle rotates in phase with the bias
field. The above arguments lead, therefore, to
\be
% \begin{split} 
% n_z &=-\cos{\vartheta}= - \frac{ m \, g }{ 2\, b\,  \mu} \, , \quad \\
% n_x &=-\sqrt{1-n^2_z} = - \sin \vartheta = - \frac{ m\, \omega^2}{ b\,  \mu } r \, , 
n_x =-\sqrt{1-n^2_z} = - \sin \vartheta = - \frac{ m\, \omega^2}{ b\,  \mu } r \, , 
\label{nr}
%\end{split}
\ee 
where  ${\vec n}= ( -\sin{\vartheta} , 0, -\cos{\vartheta})$ and $r$ being the distance
of the equilibrium point from the origin, i.e. the radius of the particle orbit in
the original frame at rest. This latter quantity can be derived by the condition of
normalization of $\v{n}$, and it results
\be
r = \frac{\mu b}{m \omega^2} \sqrt{1 - \left( \frac{m g}{2 \mu b} \right)^2 } \, .
\label{r}
\ee
The magnetic moment configuration for the dynamical
regime of this trap, and its relative position with respect to the bias field and
total field are shown in Fig. \ref{fig1}. Finally the anti-alignment condition
between $\v{n}$ and $\v{B}_0$ leads to
\be
\frac{{B_0}_z}{{B_0}_x} =
{  \frac{-2 \, b\,z +\omega }{B_{b} + b \, r}}
= \frac{1}{\tan{\vartheta}} \, .
\label{teta}
\ee
By combining this latter equation and the one derived from the Eqs. (\ref{nz})
and (\ref{nr}), we can obtain the $z$ equilibrium position.
From Eqs. (\ref{r}) and (\ref{teta}) we can observe that if $\omega$ changes its sign,
by keeping fixed all the other parameters of the trap, $\vartheta$
does not undergo any change, as can be gather from Eqs. (\ref{nz}) and (\ref{nr}), and a displacement in the $z$ position of amplitude $-\frac{\omega}{b}$ will therefore occur.

%%%%%%%%%%%%%%%%%%%%%%%%%%%%%%%%%%%%%%%%%%%%%%%%%%%%%%%%%
\subsection{The pseudo-periodic orbits} \label{ppo}

We derive in the present subsection a particular kind of orbits that are
important for the discussion we will make in the following about the results
contained in Ref. \cite{ARI}. In the case in which the bias field $B_b$ is much
greater of its rotation frequency $\omega$, a class of orbits which are almost
periodic do exist, even if the magnetic moment is not exactly anti-aligned
with total field ${\vec B}_0$. As we have recalled in the discussion about
the adiabatic approximation, whatever the initial angle between the dipole
and the field is, it will remain unchanged since the dipole performes a very
fast processional motion around the magnetic field.
As far as the orbital motion is concerned, the fast oscillation of the
dipole, that happens on shorter time scales, will be averaged out, and,
consequently, the force on the dipole will depend only on its time averaged value.
When a motion similar to that described above will take place, the
average vector $\left\langle {\vec n}  \right\rangle$ replaces the original
vector $\vec n$, in the expressions (\ref{nz})-(\ref{teta}) for the periodic
motion given in the previous section. Thus, $\left\langle {\vec n}
\right\rangle$ will be anti-aligned with respect to ${\vec B}_0$, and will
have intensity $| \left\langle {\vec n}  \right\rangle | = |\cos
\psi |$ where $\psi$ is the initial angle between the dipole and the
field. Thus, from Eqs. (\ref{nz}), (\ref{nr}) and (\ref{teta}) we deduce
\be
\begin{split}
 \left\langle n_z \right\rangle &=
	-|\cos\psi| \cos\vartheta = -\frac{mg}{2 b \mu} \, , \\
\left\langle n_x \right\rangle  &= - |\cos\psi| \sin\vartheta =
- \frac{m \omega^2}{b \mu} r \, ,
%  r&=  - \frac{b \mu}{m \omega^2} \left\langle n_x \right\rangle  =
% \frac{b \mu}{m \omega^2}|\cos\psi| \sin\vartheta \, ,
\label{nzmr}
\end{split}
\ee
that combined with the equation
\be
\frac{-2 b z + \omega}{B_b + b r} = \frac{1}{\tan \vartheta} 
\label{height}
\ee
determine the orbits height.
It is worth to pointing out that, in this case, the orbit $z$ coordinate
depends also on the dipole-field angle $\psi$ as also $\vartheta$ does, in
virtue of Eqs. (\ref{nzmr}). In other words this orbit $z$-coordinate depends
on the dipole initial conditions.

One may ask if, such an orbit, has any relation with the trapping mechanism
operating in the BEC experiments. In this respect it should be remarked that
the only classical orbits which correspond to a stationary quantum state are
periodic orbits. For these trajectories, the wave associated to the orbital motion
undergoes constructive interference as it returns in the same position in space,
after that a period has elapsed. This condition gives account for the old
Sommerfeld's quantization rule. On the contrary, in the quasi-periodic orbit,
the magnetic dipole evolution, is quite independent of the orbital evolution
and the particle returns in the same spatial position with its dipole state
differing from that of the previous passage through that point. This would
lead to destructive interference of the wave associated to the particle.
Therefore, the pseudo-periodic orbits can hardly be associated to any
stationary quantum state of the system.

%%%%%%%%%%%%%%%%%%%%%%%%%%%%%%%%%%%%%%%%%%%%%%%%%%%%%%%%%
\subsection{Oscillations in the vertical position} \label{oscillations}

In the presence of a strong adiabatic field, we can easily
describe the oscillations which take place when the initial $z$-position
of the trapped particle is different from the $z$ position given by
Eq. (\ref{height}), in the case of a pseudo-periodic trajectory with initial
angle $\psi$. Indeed, the potential energy in the case of strong adiabatic
motion is given by
\be
U = mgz + \mu |\cos \psi| \sqrt{(B_b + b r)^2 + (-2 b z + \omega)^2}
\label{potener}
\ee
to which it corresponds the force
\be
F_z = - mg - 2 b \mu \left\langle n_z \right\rangle =
-mg + 2 b \mu |\cos \psi| 
\frac{(-2bz+\omega)}{\sqrt{(B_b+br)^2+ (-2bz + \omega)^2}} \, .
\ee
Thus, when $r$ is small such as $b r$ can be neglected with respect to
$B_b$, this force depends only on $z$, and the resulting motion is an
oscillatory one around the $z$ equilibrium position derived by Eq. (\ref{height}).
This oscillating regime is important for the
discussion that will follow. In fact the $z$ equilibrium position of
a BEC cloud is experimentally evaluated just by measuring the average
value of these oscillations.

%%%%%%%%%%%%%%%%%%%%%%%%%%%%%%%%%%%%%%%%%%%%%%%%%%%%%%%%%
\section{Triaxial trap motion} \label{triaxial}
In the case of a triaxial trap it is no more possible to obtain an
exact solution for the periodic orbit as done above. Indeed, the rotational
symmetry about the vertical axis is lost and, consequently, the magnetic field
$\vec{B}_0$ in the frame rotating with the bias field, remains a time-dependent
quantity. In order to overcome this problem,
we will perform a perturbative calculation in which  we will take as a basis
the periodic orbit that takes place in this trap, for a very small
$b$ parameter.
Therefore, in Sec. \ref{zero}, we will derive such an orbit, then, by using
the result of calculation of App. \ref{small}, we will
evaluate the effect that the small inhomogeneous static magnetic field,
seen as a varying one by the dipole during its motion, produces on the
dipole orientation in space.
We shall show that in general, the dipole orientation, after that the
particle returns in its initial position in space, will be changed.

Thus, an additional rotation that brings the magnetic dipole back to its
initial state, can be obtained by a static magnetic field.
This magnetic field compensation will be obtained by a vertical shift
of the orbit in space.
Of course, such an adjustment is not needed for the symmetric trap,
because the inhomogeneous field is seen as constant one during the orbital
motion.
It is noteworthy that, such a repositioning of the orbit, will turn out to be
different for a bias field rotating in the positive or in the negative sense.
This latter point will be shown in Sec. \ref{difference}

%%%%%%%%%%%%%%%%%%%%%%%%%%%%%%%%%%%%%%%%%%%%%%%%%%%%%%%%%
\subsection{Zero order solution} \label{zero}
In this section, we derive the periodic orbit that takes place for a
very small $b$ parameter.
In order to accomplish this, we neglect the inhomogeneous field components
in the $x-y$ plane, only in the torque equation. Thus we write
$$
\dot{\v{n}} =  \v{n} \times (B_b \cos(\omega t),B_b \sin(\omega t), -b z) \, .
$$
We can assume the adiabatic approximation i.e that $\vec  n$ is anti-aligned
with respect  to the approximate magnetic field, which is given by the bias
field and by the $z$ component of the inhomogeneous field. This condition allows
us to write, in the laboratory frame, $n_x=-n_\perp \, \cos{\omega \, t}$  and
$n_y=-n_\perp \, \sin{\omega \, t}$, where $n_\perp$ indicates the intensity
of the component of the dipole unit vector in the plane $x-y$. Let $n_z$ be
the intensity of the $z$ component of the dipole unit vector. Thus, the
anti-alignment condition gives the equation
\begin{equation}
\frac{n_\perp}{n_z} = \frac{B_b}{(b z -\omega)} \, .
\label{nnbb}
\end{equation}
The forces on the particle are
\be 
\begin{split}
F_x &= - 2 \, \mu \, b \, n_\perp \,  \cos(\omega t)   \, , \\ 
F_y &= \mu \, b \, n_\perp \,   \sin(\omega t)  \, ,  \\
F_z &= -\mu \, b \, n_z  -m\, g \, ,
\label{forces}
\end{split}
\ee
by solving the Newton's equations in the $x-y$ plain, we get the time law for
the position coordinates $x$ and $y$ in the laboratory frame
\be
x (t)= {\frac{2\, \mu \, b}{ m \, \omega^2 }} \, n_\perp  
\, \cos(\omega t)  \, , \, \textrm{and} \quad
y (t)= - {\frac{\mu \, b}{ m \, \omega^2 }} \, n_\perp   \, \sin(\omega t)
\, ,
\label{elliptic}
\ee
where $n_\perp$ is still an unknown quantity.
Thus, for small $b$ the orbit takes place entirely in a plane
parallel to the $x-y$ one on an elliptic
trajectory, and is counter-rotating with respect to the bias field.
The force equation along the $z$ axis provides the angle by which $\v{n}$ 
must be tilted with respect to this axis, and the value of $n_\bot$.
Indeed, the equation $F_z = 0$ gives, from the third of Eqs. (\ref{forces}),
\be
-n_z = {\frac{ m \, g }{ \mu \, b}}:= \cos{ \vartheta} \,
, \, \textrm{and} \quad
n_\bot = \sin \vartheta \, .
\label{angle}
\ee
% The anti-alignment condition is easily established since the $x-y$ plane components
% of the inhomogeneous field are neglected.
By combining the latter equations with the one in (\ref{nnbb}), we get
\be
{  \frac{- b\,z +\omega }{ B_b}}= \frac{1}{\tan{\vartheta}} \, ,
\ee
which with (\ref{angle}) makes it possible to determines the orbit's
$z$ level.

In order to include in the torque equation the contribution given by inhomogeneous
field in the plane $x-y$, we have to compute the magnetic field $\v{b}^\prime(t)$,
in the $x-y$ plane, that the dipole  experiences along its trajectory for effect
of the inhomogeneous field.
In the laboratory frame this is given by $\v{b}^\prime(t) = (2 b x(t) , -
b y(t) )$, where $x(t)$ and $y(t)$ are given in Eq. (\ref{elliptic}).
It is convenient to refer this field respect to the system frame rotating with the bias
field, it is not difficult to find
\be
\begin{split}
%\begin{eqnarray} 
b_x^\prime &=\frac{3}{2}  {\frac{ \mu \, b^2}{ m \, \omega^2 }} \, 
n_\perp \, \cos{2\,\omega\,  t} + \frac{5}{2} {\frac{\mu \, b^2} 
{ m \, \omega^2 }} \, n_\perp  \, , 
%\non 
\\
b_y^\prime &=-\frac{3}{2} {\frac{ \mu \, b^2}{m \, \omega^2}} \, 
n_\perp   \sin{2 \omega\, t}  \, .
\label{rb}
%\end{eqnarray} 
\end{split}
\ee
Thus, the inhomogeneous field component on the dipole results to be the sum of a
constant field and a small time varying component.
Apart form the constant term which resets the value of the bias field, as
occurred also in the previous case concerning  the symmetric trap, the
sinusoidal components
produces deviations from the anti-alignment condition. In order to evaluate these
deviations, and how they can be compensated, we will solve the torque equation under the
presence of a constant field complemented with two small oscillating components.

%%%%%%%%%%%%%%%%%%%%%%%%%%%%%%%%%%%%%%%%%%%%%%%%%%%%%%%%%
\subsection{Difference of height between $\omega$ and -$\omega$} \label{difference}

In the present section we will apply the results found in appendix \ref{zero},
making explicit Eq.~(\ref{NAD}), in the case of the motion in the triaxial
trap.
Before making these equations explicit for the case of the triaxial trap,
we evaluate the order of magnitude of the deviation from the
adiabatic approximation, a fair indication of this being the angle between
$\v{n}_0$ and $\v{B}_0$. From Eq. (\ref{NAD}) is clear that this quantity
increases as $\v{B}_0$ gets closer to $\omega$.
However, in the usual experimental setups, this condition is carefully avoided
because it is known to lead to instabilities in the trap. Indeed,
reasonable experimental values for the quantities that appear in (\ref{NAD}) are
$B_0 \approx B_b$ of the order of $10^4$ and $\omega = 2 \pi$ in our units.
Thus, if we assume the condition in which $\omega \ll B_0$, the second term of
Eq. (\ref{NAD}) dominates over the first one. Its intensity is of the
order of $\frac{b^2_0}{B_0^2}$ with $b_0$ standing for the order of magnitude of
$|\v{b}_1|$ and $|\v{b}_2|$ of (\ref{B0b1b2}). Now, for typical experimental parameters,
$b_0 = \frac{3}{2} \frac{\mu b^2}{m \omega^2} n_\bot \approx
\frac{\mu b^2}{m \omega^2} = 0.92$ (here we have used $b=10.$).
Thus the amplitude of the corrections due to the second term in Eq. (\ref{NAD})
results of the order of $10^{-8}$.
The first term of the same equation, is instead of the order of
$\frac{\omega}{B_0} \cdot  \frac{b^2_0}{B_0^2}$, and then, it results to be smaller.
However, its dependence on $\omega$ is important and it may perhaps become
more relevant for trap where the adiabatic condition is not so strongly
enforced, as in the case of the experimental setup being considered here.
We will derive explicitly the shift entailed from this term in the following
of the section. It must be noticed
also that, given the relative size of the two terms in the above equation,
the adiabatic approximation is broken earlier than the effect of the reversed
rotation can be observed.

\begin{figure}[ht]
\includegraphics[angle=270,scale=0.38]{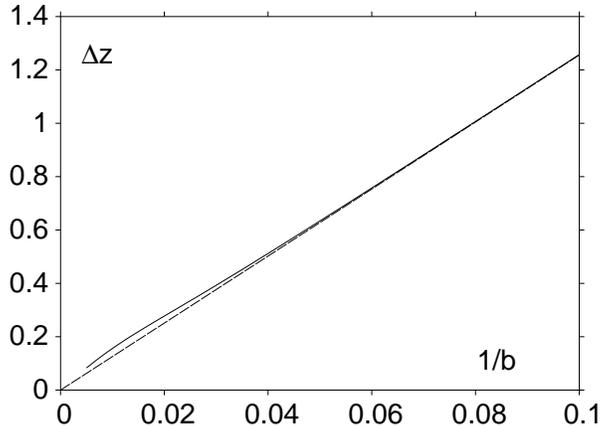}
\caption{ The shift obtained with the theoretical calculations (continuous 
line) and the zero order compensation  shift (dotted line).
\label{fig2}}
\end{figure}

In order to use the result of appendix \ref{zero}, we adopt a frame rotating
with the bias field and we set
\be 
b_0=\frac{3}{2}  {\frac{\mu \, b^2}{m \, \omega^2}}\, n_\perp \, .
\label{pertb}
\ee
The only components of the fields $\v{b}_1$ and $\v{b}_2$ of Eq. (\ref{B0b1b2}),
different from zero in this frame are $b_{1,x}=b_0$
and $b_{2,y}=-b_0$, as they can be read from Eq.~(\ref{rb}).
Moreover we have
$\v{B}_0= (B_b + \frac{5}{3} b_0,\,  0,\,  -b\, z + \, \omega)$
and  $\v{n}_0=(-\sin{\vartheta}, \, 0., \, -\cos{\vartheta})$,
where the angle $\vartheta$ is that one corresponding to the unperturbed
motion and given by Eq.~(\ref{angle}).
Since initially $\v{n}_0$ and $\v{B}_0$ are opposite to each other,
Eq.~(\ref{NAD}) prescribes how one of them should be modified.
However since $\v{n}_0$, and consequently the angle $\vartheta$ is
already determined by the equilibrium equation along the $z$ axis,
only $\v{B}_0$ can be modified. This modification amounts to a
small time-independent contribution obtained, in the specific case
of the triaxial trap, by a shift of the whole orbit in the $z$
direction. In order to calculate the angle between $\v{n}_0$ and
$\v{B}_0$ we note that the only component of $\v{n}_0 \times \v{B}_0$
different form zero is given by
\be 
(\v{n}_0 \times \v{B}_0 )_y =\frac{1}{2} \left(\frac{b_0}{\Delta}
\right)^2 
( -2\,\omega \, \sin{\vartheta} + B_0 \sin{\vartheta} \, \cos{\vartheta})\, ,
\ee
where one should remind that the angular frequency of the oscillating field
component is $2 \omega$ instead of $\omega$ (Cf. Eqs. (\ref{rb}) and (\ref{B0b1b2})).
The above equation makes evident that $\v{n}_0$ still lies in the
$x-z$ plane, and that the tilting angle of $\v{B}_0$ with respect to $\v{n}_0$,
whose sign is determined by the $y$ axis, is now given by
\be 
\Delta \vartheta= -\frac{1}{2} \left(\frac{b_0}{\Delta}
\right)^2 \left( -\frac{2\,\omega}{B_0}  \, \sin{\vartheta} + 
\sin{\vartheta} \, \cos{\vartheta}\right) \, ,
\ee
where $B_0$ is the intensity of the total magnetic fields and
$\Delta^2= B_0^2 -
4\,\omega^2$.
The new component ${B_0}_{,z}$ must now satisfy the condition
$\frac{{B_0}_{,z}}{B_b + \frac{5}{2} b_0}= \frac{1}{\tan{(\vartheta
+\Delta \vartheta)}}$. By using
the approximation  $\frac{1}{\tan{(\vartheta +\Delta \vartheta)}}
= \frac{1}{\tan{\vartheta}} - \frac{\Delta \vartheta} {\sin{\vartheta}^2}$ 
we obtain
\be
z=z_0 + \frac{1}{2}\left( \frac{b_0}
{\Delta} \right)^2
\left(z_0 + \frac{\omega}{b} \right) \, ,
\ee
where $z_0$ is the $z$ position of the orbit in the zero approximation.
From this expression it results that the shift in $z$, caused by the
change $\omega \rightarrow -\omega$, for which $\Delta z_0 = -\frac{2
\omega}{b}$ and $\Delta \omega = - 2 \omega$, is given by
\be 
\Delta z =  -\frac{2 \,\omega}{b} \left[ 1 +  
\left(\frac{b_0}{\Delta}\right)^2 \right] \, .
\label{displ}
\ee
This expression depends on $b$ but, considered the small size
of the perturbing field $b_0=\frac{3}{2}\frac{\mu \, b^2}{m \, \omega^2}\,
\sin{\vartheta}$, it remains very small anyway
even if $B_0$ becomes close to the bias field frequency $\omega$ where,
on the other hand, the perturbative calculation presented here would no
longer apply. It is important to point out that the correction found above
depends, quite naturally, on the ratio between the strength of the
average inhomogeneous field experienced by the particle during its motion
and the total static field $B_0$. Then, it is approximatively given by the
ration between $b_0$, and $B_b$.
The predicted shift is shown in Fig.~\ref{fig2}
and compared with the zero order compensation shift
$\Delta z_0 = - \frac{2 \omega}{b}$.
As it is expected the difference between the two is really a very small one
even for the parameter used in a real trap of Fig.~\ref{fig2}.
% On the other hand for parameters used in a real trap the difference would
% not either be visible because, as shown above, the relative correction in
% Eq.~(\ref{displ}) is of the order of $10^{-8}$. Although the perturbative
% calculation does not really apply for big values of $b$ the expression in
% Eq.~(\ref{displ}) yields a zero value as $b$ increases in virtue of the
% static term in Eq. (\ref{rb}).

%%%%%%%%%%%%%%%%%%%%%%%%%%%%%%%%%%%%%%%%%%%%%%%%%%%%%%%%%
\subsection{Non-adiabatic regime and bi-stability of the trap}

In order to compare the above theoretical derivation with the
results obtained by numerically integrating the equations of motion (\ref{SE}),
we have determined, by numeric computations,
the periodic orbits for this system. The method, which is
customarily used in investigations of this kind, assumes that
a periodic, and then closed, orbit, named reference orbit, is known for
certain values of the system parameters.
A small change of one of these parameters opens in space, by a slight amount,
the reference orbit when evolved over a period of the time-dependent force.
Thus, a small variation of the initial condition could recover
the periodicity, and close the orbit in space again. A linear map
connecting initial and final values, for all orbits that lay close to
the reference one, can thus be determined by performing evolutions of the
system with initial conditions very close to those of the reference orbit.
This map does not appreciably change as the parameters
undergo small changes. Thus the new periodic orbit can be found as the fixed
point of this map by inverting numerically the related matrix.
In our case the initial reference orbit is that
corresponding to a small value of the parameter $b$. Starting from this
orbit, we can determine numerically the periodic orbit of the system for
any given parameter value $b^*$ by means of many small
changes of the $b$ parameter, until the final value $b^*$
is reached.

First of all, we have used this numerical procedure to compare
the shift in the $z$ position of the orbit with that derived
in the previous section. More precisely we have numerically calculated
averaged value of this quantity, and compared it with Eq.~(\ref{displ}).
\begin{figure}[ht]
\includegraphics[angle=270,scale=0.38]{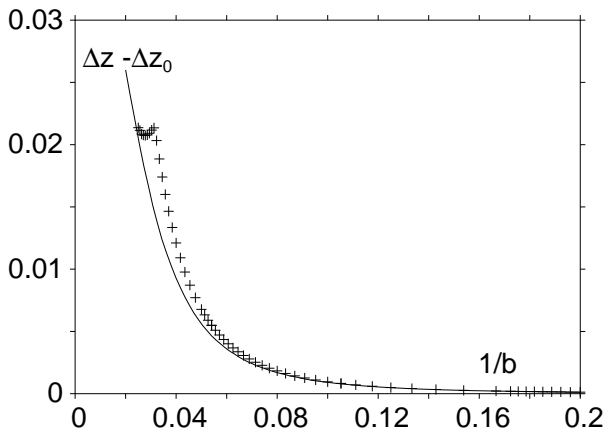}
\caption{ Comparison between the theoretical shift (continuous line)
and the numerical one (crosses).
In order to make more evident this comparison the
zero order shift ${\Delta z}_0=\frac{2 \omega}{b}$
has been subtracted to both of these two quantities.
At small values of the $\frac{1}{b}$ the regularity of this dependence
is somewhat broken as shown by the little terminating wiggle.
This corresponds also to orbits which are on the point of becoming unstable.
The parameters used for this figure are $B_{\rm r}=50.$, $\frac{\mu}{m}$=0.036,
$g=0.1$.
\label{fig3}}
\end{figure}
As reported in Fig.~\ref{fig3}, we observe that the agreement is
quite good even for values of $b$ in a more wide range of those commonly
encountered  in experimental setups. The calculated shift
is extremely small such as being, at the present, out of the reach of
experimental verification, and certainly far away from the values observed
in the experiment of Ref. \cite{ARI}.

By using the numerical routines that we have designed for the analysis
of the periodic orbits, we have performed a coarse examination of the
unstable dynamical regimes of the trap.
\begin{figure}[!t]
\includegraphics[angle=270,scale=0.38]{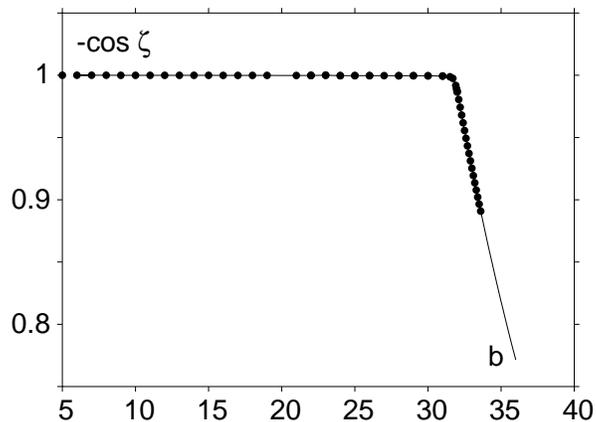}
\caption{ The cosine of the angle between the magnetic field and the
magnetic dipole at time $t=0$ of the periodic orbit, as a function
of the $b$ parameter for both $\sigma^+$ (continuous line) and $\sigma_-$
(big dots) polarization of the bias field.
\label{fig4}}
\end{figure}
\begin{figure}[!t]
\includegraphics[angle=270,scale=0.38]{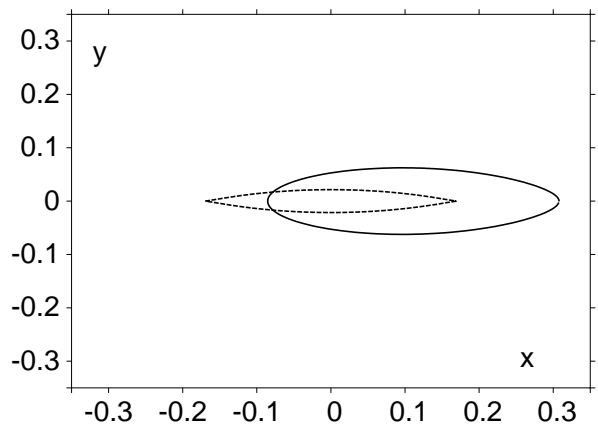}
\caption{ Two  bistable orbits corresponding  to trap values of
$b=10.$, $B_{\rm r}=11.36$ and $\sigma^+$ (anti-clockwise) bias field.
\label{fig5} }
\end{figure}
It is found that signatures of instability
appears, quite naturally, as the intensity of the inhomogeneous field
experienced by the particle becomes of the same order than that of the bias
field. For these values the peculiar scenario leading to unstable behavior
with orbit bifurcations, orbit bi-stability and eventually
chaotic behavior will emerge. However the instability is always connected,
at least for the range of the parameters observed here, to the loss of
the system adiabatic condition. We have numerically determined the angle
between the effective field $\v{B}_{0}$ and the dipole vector
at time $t=0$, i.e. when the bias field crosses the positive $x$ axis.
This is shown in Fig.~\ref{fig4} for both directions of bias field
rotation. As it can be seen here the adiabatic approximation is quite
resilient even if the trap parameters are rather loose, in terms  of the
adiabatic condition enforcement, with respect to those normally used in a
TOP trap. For the same range of values of the $b$
parameter where an appreciable dis-alignment occurs we have observed
a bi-stability  regime. The two orbits involved in this regime are
shown in Fig. \ref{fig5}.
The analysis of the unstable regimes of this system would require a
systematic study that is out of the purpose of the present work.

%%%%%%%%%%%%%%%%%%%%%%%%%%%%%%%%%%%%%%%%%%%%%%%%%%%%%%%%%
\section{Comparison with the experimental results} \label{comments}

There is an evident discrepancy between the results obtained here
and the experimental ones reported in Ref. \cite{ARI}.
One of the hypothesis that
we can follow to deal with this situation is to assume that a
classical model is not an adequate one to describe the dynamics
of an atom in the triaxial TOP trap and that a quantum calculation
is in order.
% To fully support this point of view we should also ponder about the
% numerical simulations carried out to support the experimental
% results. Since the details about these numerical simulations
% have never been reported in the literature and from the original
% work it is not possible gain sufficient insight in this respect,
% we can offer here only reasonable and quite plausible hypothesis
% on how they were carried out. As a first remark we notice that
% the correspondence between quantum states and periodic orbits has
% never been called upon there. Therefore it is quite likely that they
% were performed on the basis of the pseudo-periodic orbits that we have
% discussen in Section \ref{ppo}.
Before to assume the above hypothesis as the final one, let us to try
an alternative analysis that take in account of a procedure for the
bias-field reversal, similar to that one actually performed in the
experiment of \cite{ARI}.
In the reasoning which follow we will refer to a symmetric trap
considered that the perturbation to the magnetic dipole motion
produced by the inhomogeneous field of the triaxial trap are really
negligible for the experimental parameters. Thus the effect that we
are going to describe will occur indifferently for the symmetric
and triaxial trap. Let us suppose that the system moves actually on
a quasi-periodic orbit with an angle $\psi$ between dipole and
field. We remind that this angle does determine the $z$ position of
the orbit according to Eq. (\ref{nzmr}) and Eq. (\ref{height}).
Now, if the sense of rotation of the bias field is abruptly changed
the system will experience the oscillations along the $z$ axis
discussed in Section \ref{oscillations}, because the $z$ coordinate
corresponding to the new pseudo-periodic orbit will be different
from the one in which the system is actually in. In the experiment,
the averaging of these oscillations is assumed as the $z$ coordinate
of the BEC stationary state corresponding to the new bias field
rotation sense. On the other hands, the classical equations will lead
to a shift that coincide with a zero order compensation shift only
if the angle $\psi$, and by consequence the equilibrium $\vartheta$,
does not change. On the contrary, if this does not occur an anomalous
contribution should be added to the standard compensation shift.
Indeed, this would be the case if we assume that during the switching
between the two rotation senses the dipole orientation in space does
not change. If this is the case, the angle $\psi$ will change uniquely as a
result of the change of the direction of $\v{B}_0$ in space, as a
consequence of the change of its $z$ projection by an amount of
$\Delta {\v{B}_0}{_{,z}} = \Delta \omega = - 2 \omega$. In appendix
\ref{appendix}, we have calculated the shift obtained under these
conditions, assuming that both the magnetic dipole and the field lies
in the plane $x-y$, and that $\psi$ is the initial angle between the two.
This latter quantity can, of course, assume positive or negative
values. After a bit of algebra, reported in appendix \ref{appendix}
we have found
\be
\Delta z = - \frac{2 \omega}{b} \left( 1 + \frac{m g}{2 b \mu}
\frac{1}{\sin \vartheta} \frac{\sin \psi}{\cos^2\psi} \right) \, .
\label{Deltaz}
\ee
This expression evidences how by choosing appropriately the angle
$\psi$, the trend of the experimental data, according to which
the deviation increases as $b$ approaches its minimum value,
could be reproduced. We notice that in this case, also $\vartheta$
approaches the zero value, thus amplifying even further the
expected effect. The above expression indicates that the anomalous
shift effect disappear in absence of gravity, a particularity which
coincides also with the conclusions
% of the numerical simulations
% under discussion as
reported in the experimental work. Here we are
presuming that the switching between positive
and negative $\omega$ does occur so fast that the dipole has no
time to adapt to the new field $\v{B}_0$, present after the switch
has taken place. Although this is rather unlikely to occur, an alternative
mechanism wouldn't make sense. In fact,
it would be possible to imagine that a loss of the adiabaticity could take place
during the switching process, if at certain point the intensity of
the magnetic field would get close to zero, in such a way that the angle
between the field and the dipole emerges out the switching process
changed. This, of course, would depend on how technically this switching
is carried out. If this would be the case, the new oscillation point would
be displaced by a quantity that contains a further anomalous
contribution. However, if it would be so, the anomalous $\Delta z_a$
would be only a fictitious one being dependent just on how the
switching process is carried out. Moreover the question of why the
system selects a particular value of $\psi \neq 0$ and, in addition,
of how a pseudo-periodic orbit can be put in correspondence with a
quantum state, would still remain open ones.
% Possible, although quite
% unsatisfactory, answers could be that a pseudo-periodic orbit
% corresponds to a superpostion of two quantum states having spin
% parallel to the field $S_\|= 2$ and $S_\|= 1$, which would explain
% why the average dipole is not exactly anti-aligned with the field.
% However this should be excluded if we suppose to deal with a true
% BEC cloud. Indeed our attitude here is to trust the experimental
% results.
In this respect we believe that although our arguments,
based on periodic orbits, are particular sound it might still be
possible that this is not jet sufficient to properly describe quantum
stationary states of the BEC cloud and quantum calculation should
be in order. In any case further evidences in support of some
of the possibilities expressed here should also be searched for.

%%%%%%%%%%%%%%%%%%%%%%%%%%%%%%%%%%%%%%%%%%%%%%%%%%%%%%%%%
\section{Conclusion}
In this work, we have used the classical equations of motion to investigate
some properties of the magnetic confinement of a neutral
atom in the TOP trap. Our initial purpose was that of understanding
the origin of an anomalous shift in the vertical position of a BEC
cloud that has been observed experimentally in Ref. \cite{ARI}. We have
calculated analytically this shift and found that this is not in
agreement with these measurements. In spite of this, our calculations
show that, in principle, there is no reason for this shift being limited
only to the standard compensation shift $\frac{2 \omega }{b}$, as
it is the case in the symmetrical trap. We have shown that an anomalous
shift exists in any case in the triaxial trap. In fact, the anomalous
shift calculated here is produced by the interplay between the particle
space dynamics, and that of its magnetic moment driven by the torque
equation. However its magnitude results to be extremely small for the
parameters used in the normal trap.
On the contrary, if such a shift
were to exist in line with the calculations made in \cite{ARI}, of
which account has been given here, it would be independent of the kind of
trap used, i.e. symmetrical or triaxial one, and would depend basically
by the peculiarity of the switching precess.
In addition we have observed a bi-stability regime in the range of values
of the b parameter where the adiabatic regime is lost. The systematic study
of this bi-stability regime and of the consequent unstable-dynamics, deserves
a further investigation.

In conclusion, the general feeling is that, the approach based on the
classical model, leaves so many unanswered questions, that we have the only
possible alternative to explain the experimental results by means
a quantum mechanical calculation.
In fact, in the quantum mechanical approach, the periodicity
condition of both the space and the internal
spin variables, could be much more stringent than in the classical case
thus amplifying an effect that appears otherwise to be quite a small one.

\appendix

%%%%%%%%%%%%%%%%%%%%%%%%%%%%%%%%%%%%%%%%%%%%%%%%%%%%%%%%%
% \subsection{Dynamics of the small oscillating components}
\section{Dynamics of the small oscillating components} \label{small}
In the present appendix we consider a dipole moving in a magnetic field
given by a static component with added small sinusoidal components.
In particular, the present section is aimed at
derive the small deviations from the anti-alignment, due to the effect
of the sinusoidal components like the ones of Eqs. (\ref{rb}).
Let the total field acting on the magnetic dipole be
\be 
\v{B}_0 + \v{b}_1 \, \cos{\omega \, t} + \v{b}_2 \, \sin{\omega \, t} \, ,
\label{B0b1b2}
\ee
and let us suppose that
$|\v{b}_1|$ and $|\v{b}_2|$ are very small with respect to $|\v{B}_0|$,
which is a time-independent vector. We look for a periodic solution of the form
\be 
\v{n} = \v{n}_0 + \v{n}_1 \, \cos{\omega \,t} +\v{n}_2 \, \sin{\omega \, t}
\label{newn}
\ee
where $\v{n}_1$ and $\v{n}_2$ are small deviations from the anti-alignment.
Here $\v{n}_0$ is the zero order solution
which is anti-aligned with respect to the field $\v{B}_0$.
By inserting the above expressions
in the torque equation for the dipole unit vector, and by neglecting the higher
order terms coming from both the product between the small rotating components of the
magnetic moment and the small rotating components of the magnetic field,
one obtains the following equations
\be
\begin{split}
-\omega \, \v{n}_1 &= \v{n}_2 \times \v{B}_0 + \v{n}_0 \times \v{b}_2 
\, , \\
\omega \, \v{n}_2 &= \v{n}_1 \times \v{B}_0 + \v{n}_0 \times \v{b}_1
\, .
\label{venti}
\end{split}
\ee
These equations can be solved for $\v{n}_1$ and $\v{n}_2$, thus we obtain \cite{note}
\be 
\begin{split}
 \v{n}_1 &= {\frac{ \omega \, \v{n}_0 \times \v{b}_2 - B_0 \, 
(\v{b}_1)_\perp}
 { B_0^2  - \omega^2} } \, , \\
 \v{n}_2 &= {\frac{ -\omega \, \v{n}_0 \times \v{b}_1 - B_0 \, 
(\v{b}_2)_\perp }{ B_0^2  - \omega^2} } \, .
\label{n1n2}
\end{split}
\ee
Here the subscript $\bot$ means perpendicular to the static magnetic field.
To the smallest order in the oscillating field amplitude, the constant term
in the torque equations yields
\be 
\v{n}_0 \times \v{B}_0 + \frac{1}{2}\,  ( \v{n}_1 \times \v{b}_1 
+ \v{n}_2 \times 
\v{b}_2 ) = 0 \, .
\ee
This expression shows that the anti-alignment condition between $\v{n}_0$ and $\v{B}_0$
can no longer be satisfied.
Thus we calculate the last two terms of the above equation by using the Eqs. (\ref{n1n2}),
after a bit of algebra we obtain
\bea
\lefteqn{\v{n}_0 \times \v{B}_0 =}&& \hspace{15truemm} 
\omega \,{ \frac{(\v{b}_2 \cdot \v{n}_0)\, \v{b}_1-   
 (\v{b}_1 \cdot \v{n}_0 ) \, \v{b}_2}{2 \, ( B_0^2 - \omega^2)}} 
 +
 \non 
\\ 
& &  {\frac{\{ (\v{b}_1 \cdot \v{B}_0)\, 
(\v{n}_0 \times \v{b}_1) +   
 (\v{b}_2 \cdot \v{B}_0 ) \, (\v{n}_0 \times \v{b}_2 )\}}{2 \,
( B_0^2 - \omega^2)}} \, .
\label{NAD}
\eea 
This equation indicates how $\v{n}_0$ or $\v{B}_0$ must be modified in order to satisfy
the torque equation.

%%%%%%%%%%%%%%%%%%%%%%%%%%%%%%%%%%%%%%%%%%%%%%%%%%%%%%%%%%%%%%%%%%%%%%%%%%%%%%%%%%%%%%

\section{} \label{appendix}
We assume that the dipole and the magnetic field $\v{B}_0$ lies both in the
$x-y$ plane of the rotating frame and that the system moves in a quasi-periodic
orbit. Let $\vartheta$ be the angle between the $z$ axis and the field and
$\psi$ the angle formed by the dipole with the field, both measured with reference
to the $y$ axis rotations. It is not difficult to calculate the change in the
vertical position of the orbit that occurs as a consequence of a change of the
dipole-field angle $\psi$ and of the bias pulsation frequency $\omega$. Indeed
from Eq. (\ref{nzmr}), noting that $|\cos\psi| = - \cos\psi$ for
$\psi > \frac{\pi}{2}$, we obtain
\be
\Delta \vartheta = \frac{mg}{2 b \mu} \frac{\sin\psi}{\cos^2\psi}
\frac{1}{\sin\vartheta} \Delta \psi \, ,
\label{Aeq1}
\ee
which is the change that the pseudo-periodic orbit angle $\vartheta$
undergoes. Thus from Eq. (\ref{height}) we obtain that $\Delta z$ and
$\Delta \omega$ are correlated as follows
\be
- 2 b \Delta z + \Delta \omega = - B_b \frac{\Delta \vartheta}{\sin^2\vartheta}
= - \frac{B_b}{\sin^2\vartheta} \frac{mg}{2 b \mu} \frac{\sin \psi}{\cos^2\psi}
\frac{1}{\sin\vartheta} \Delta \psi \, ,
\label{Aeq2}
\ee
where the small term corresponding to the inhomogeneous field has been neglected
with respect to $B_b$. On the other hand the change in the field direction, as a
result of small changes of the $z$ component is simply given by
\be
-\frac{\Delta \vartheta}{\sin^2\vartheta} = \frac{\Delta \omega}{B_b} \, ,
\label{Aeq3}
\ee
where the fact that the vertical position of the orbit is unmodified during
the phase that reverts the rotation of the bias field has been taken into
account. A change in the field direction reflects in a change of the
dipole-field angle and, assuming that in the mean time the dipole angle has
remained fixed, we have $\Delta \psi = - \Delta \vartheta$. Thus
\be
\Delta z = - \frac{2 \omega}{b} \left( 1 + \frac{mg}{2 b \mu}
\frac{1}{\sin \vartheta} \frac{\sin \psi}{\cos^2 \psi} \right) \, .
\label{Aeq4}
\ee
As a consequence of the instantaneous modification of the bias field rotation,
the system performs oscillations around a position which is displaced by a
quantity $\Delta z$ with respect to the initial position of the quasi-periodic
orbit.

\begin{acknowledgments}
R. F. thanks B. Z., a right man, a good friend and a brilliant physicist.
Bruno Zambon died on October 15 2006 in Pisa, before the publication
of the present paper.

\end{acknowledgments}

\vfill 

%%%%%%%%%%%%%%%%%%%%%%%%%%%%%%%%%%%%%%%%%%%%%%%%%%%%%%%%%
%%%%%%%%%%%%%%%%%%%%%%%%%%%%%%%%%%%%%%%%%%%%%%%%%%%%%%%%%
%%%%%%%%%%%%%%%%%%%%%%%%%%%%%%%%%%%%%%%%%%%%%%%%%%%%%%%%%

%%%%%%%%%%%%%%%%%%%%%%%%%%%%%%%%%%%%%%%%%%%%%%%%%%%%%%%%%


\begin{thebibliography}{33}
\bibitem{note} Here the condition
   $\v{n}_1\cdot \v{B}_0=0$ and $\v{n}_2\cdot \v{B}_0=0$, as it results from the
   Eqs. (\ref{venti}),  has been used.
\end{thebibliography}
\end{document}